\documentclass{ws-procs9x6}

\usepackage{color}

\begin{document}

\title{CHARM AND CHARMONIUM IN THE QUARK-GLUON PLASMA}

\author{R.~RAPP$^*$, D.~CABRERA and H.~VAN HEES}

\address{Cyclotron Institute and Physics Department, Texas A\&M University,\\
College Station, Texas 77843-3366, U.S.A.\\
$^*$E-mail: rapp@comp.tamu.edu\\
}

\begin{abstract}
  In the first part of the talk, we briefly review the problem of
  parton-energy loss and thermalization at the Relativistic Heavy-Ion
  Collider and discuss how heavy quarks (charm and bottom) can help to
  resolve the existing experimental and theoretical puzzles. The second
  part of the talk is devoted to the properties of heavy quarkonia in
  the (strongly interacting) Quark-Gluon Plasma (sQGP) and their
  consequences for observables in heavy-ion collisions.
\end{abstract}


\bodymatter

\section{Introduction}
\label{sec_intro}

Experiments at the Relativistic Heavy-Ion Collider (RHIC) suggest that
the matter created in semi-/central Au-Au collisions at
$\sqrt{s_{NN}}$$=$$200GeV$ constitutes an equilibrated, strongly
interacting medium at high energy density, well above the critical one
at the expected phase boundary, $\epsilon_c$$\simeq$~1\,GeV/fm$^{3}$
(cf.~the experimental assessment papers\cite{phenix05,star05} and
references therein).  This conclusion is essentially based on three
evidences: (i) at low transverse momenta, $p_T$$\le$2-3\,GeV (comprising
$\sim$99\% of the produced hadrons), ideal relativistic hydrodynamics
describes well the single-particle spectra and their azimuthal
asymmetry, $v_2(p_T)$; the underlying collective expansion follows from
kinetic equilibration times of around $\tau_0$$\simeq$0.5\,fm/$c$, implying
rapid (local) thermalization of the system at initial energy densities
of about $\epsilon_0$$\simeq$30\,GeV/fm$^3$.  (ii) at high momenta,
$p_T$$\ge$5-6\,GeV, the production of hadrons is suppressed by a factor
of 4-5 relative to elementary $p$-$p$ (or d-Au) collisions. This has
been interpreted as energy loss of energetic partons traveling through
an almost opaque high gluon-density medium. (iii) at intermediate
momenta, $p_T$$\simeq$2-6GeV, surprisingly large baryon-to-meson ratios,
as well as an empirical scaling of the hadron $v_2$ according to its
constituent quark content, has been interpreted as quark coalescence as
the prevalent hadronization mechanism in this regime, and thus as
partonic degrees of freedom being active. The microscopic mechanisms
underlying thermalization and energy loss, however, are not well
understood yet. In fact, the jet quenching of light partons is so strong
that high-$p_T$ particle emission appears to be mostly limited to the
surface of the fireball, rendering a precise determination of the
corresponding transport coefficient (and maximal energy density)
difficult. While earlier approaches assumed the dominance of radiative
energy loss via medium-induced gluon radiation (which prevails in the
limit of high jet energies) cf.~\cite{Gyulassy:2003mc}, it has been
realized subsequently that elastic (2$\leftrightarrow$2) scattering
processes cannot be neglected at currently accessible
$p_T$$\le$10-15\,GeV at RHIC.  Moreover, none of the evidences (i)-(iii)
is directly connected to the fundamental properties distinguishing the
Quark-Gluon Plasma (QGP) from hadronic matter, namely the deconfinement
of color charges and the restoration of the spontaneously broken chiral
symmetry (which is believed to generate most of the visible mass in the
universe).

Heavy-quark (HQ) observables are expected to provide new insights into
the aforementioned problems (with the exception of chiral symmetry
restoration). On the one hand, charm ($c$) and bottom ($b$) quarks, due
to their relatively large mass, should suffer less energy loss and
undergo delayed thermalization when traversing the QGP, and thus be more
sensitive to the interactions with thermal partons than light quarks and
gluons. Therefore, it came as a surprise when the measurement of
single-electron ($e^\pm$) spectra associated with the semileptonic decay
of open-charm (and -bottom) hadrons showed a factor 4-5
suppression\cite{phenix-raa,star-raa}, very comparable to what has been
found for pions. In addition, the observed values for the $e^\pm$
elliptic flow, $v_2^e(p_T)$, reach up to 10\% at $p_T$$\simeq$2\,GeV,
indicating the build-up of substantial \emph{early} collectivity of $c$
quarks.  These observations reinforced the implementation of elastic
energy-loss processes into the theoretical description of the
spectra\cite{Hees:2004,Moore:2004,Mustafa:2005,Hees:2005,Wicks:2005gt}.
On the other hand, HQ bound states (quarkonia), due to their small
size/large binding, are suitable probes of a surrounding medium of
sufficiently high density. E.g., for a typical (ground-state) charmonium
of size $r$$=$0.25\,fm, the relevant parton density at which significant
modifications are to be expected is $n $$\approx$$3/(4\pi
r^3)$$\approx$$ 10$-$20$\,fm$^{-3}$, which for an ideal QGP (with
$N_f$$=$2.5 massless flavors) translates into a temperature of
$T$$\approx$270\,MeV$\approx$$ 1.5T_c$. Indeed, recent lattice QCD
(lQCD) calculations indicate that $J/\psi$ and $\eta_c$ states survive
in a QGP up to $\sim$2$T_c$. The theoretical challenge is then to
disentangle and quantify (a) medium modifications of the binding
potential (e.g., color screening), (b) parton-induced dissociation
reactions, and (c) in-medium changes of the HQ mass which affects both
the bound-state mass and its decay threshold, as well as to identify and
establish connections to observables in heavy-ion collisions.
 
In the first part of the talk (Sec.~\ref{sec_open}), we evaluate HQ
diffusion and thermalization in the QGP employing both perturbative and
\emph{nonperturbative} elastic interactions. This problem is
particularly suited to a Fokker-Planck approach for Brownian motion in a
thermal background\cite{Svet88}, which allows to describe both the
quasi-thermal and kinetic regime, and its transition. A Langevin
simulation for $c$ and $b$ quarks in an expanding QGP at RHIC is
supplemented with coalescence (and fragmentation) at $T_c$, thus
implementing all of the three main features (i)-(iii) of the initial
RHIC data. In the second part of the talk (Sec.~\ref{sec_hidden}), we
discuss in-medium properties of heavy quarkonia, including information
from lQCD. We outline a $T$-matrix approach in which HQ potentials from
lQCD are used as input to simultaneously describe bound and scattering
states. This enables a comprehensive evaluation of euclidean correlation
functions, which in turn can be checked against rather accurate results
from lQCD.  We briefly discuss how in-medium quarkonium properties
reflect themselves in observables in ultrarelativistic heavy-ion
collisions. Sec.~\ref{sec_concl} contains our conclusions.

\section{Open Charm and Bottom in the QGP}
\label{sec_open}
\begin{figure}[!t]
\begin{minipage}{5.5cm}
\vspace{-0.2cm}
\psfig{file=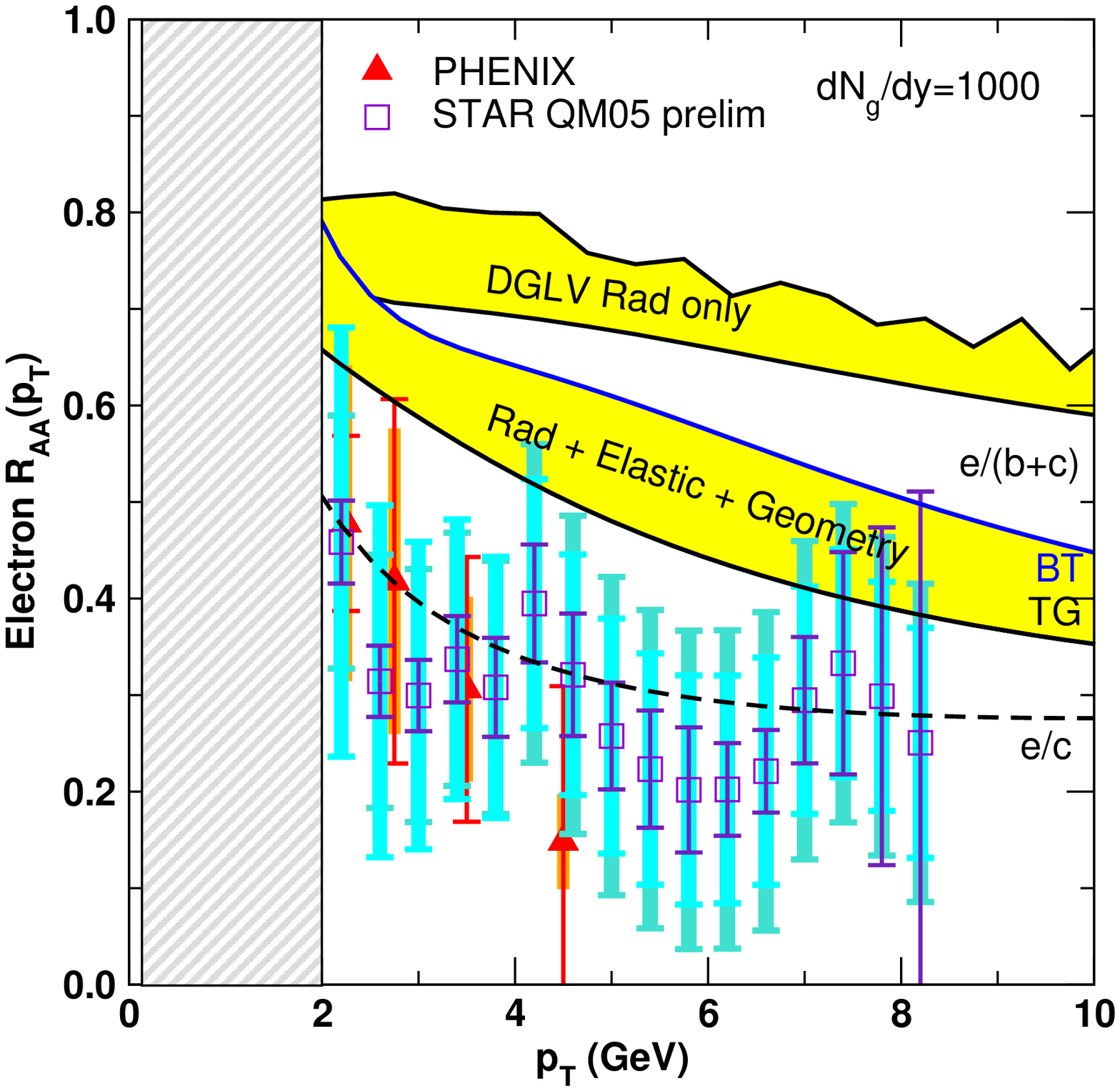,width=0.95\linewidth}
\end{minipage}
\hspace{0.2cm}
\begin{minipage}{5.5cm}
\psfig{file=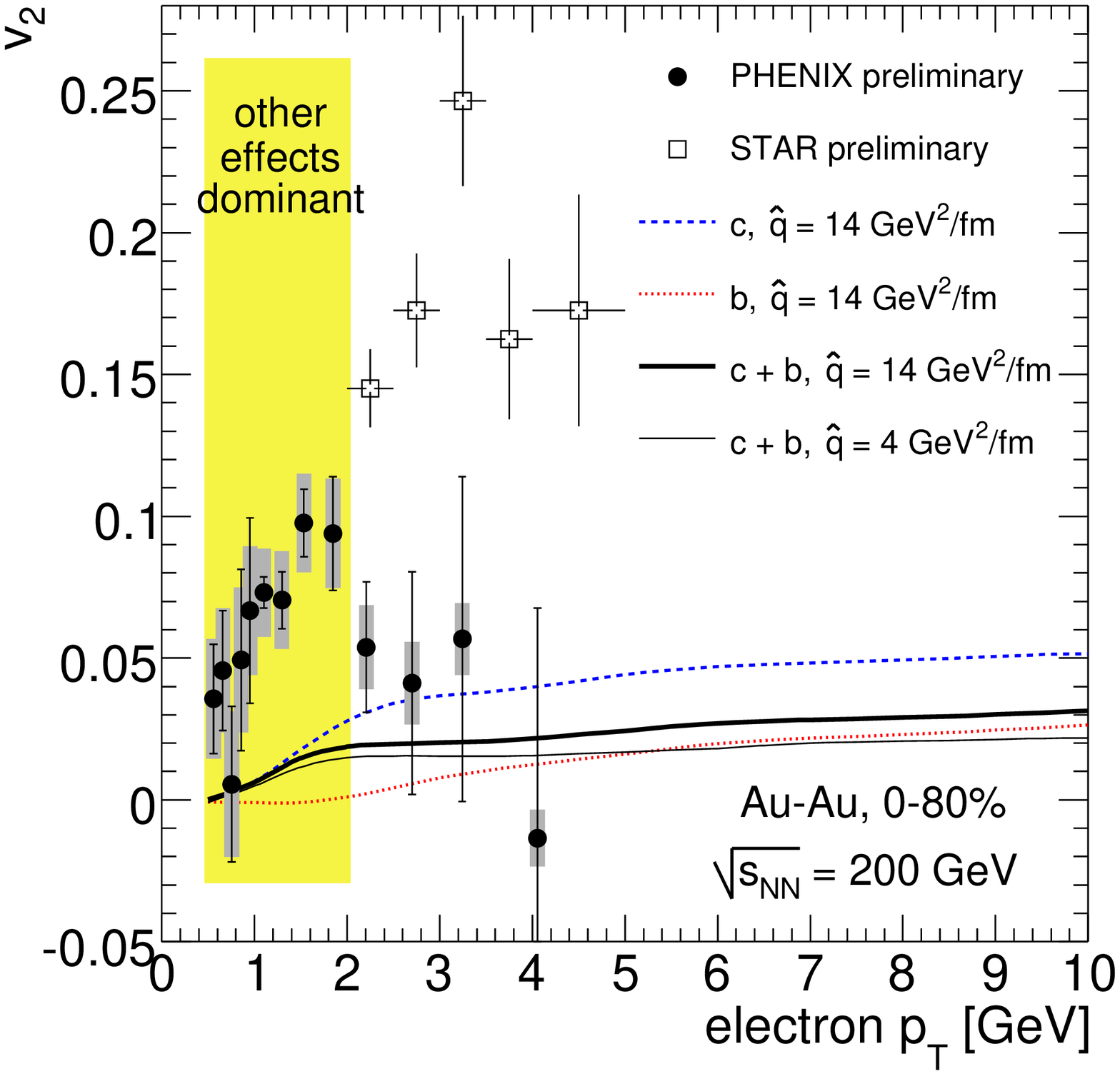,width=0.98\linewidth}
\end{minipage}
\caption{Perturbative QCD energy-loss calculations for $e^\pm$ 
  spectra from semileptonic heavy-meson decays.  Left panel:
  nuclear suppression factor\protect\cite{Wicks:2005gt}, right panel:
  elliptic flow\protect\cite{Armesto:2005mz}.}
\label{fig_pqcd}
\end{figure}

Recent calculations of hadronic spectral functions in a QGP, both within
lQCD and lQCD-based potential models, suggest that resonance (or bound)
states persist up to $2T_c$, for both heavy- ($Q\bar Q$) and
light-quark ($q\bar q$) systems. We conjectured\cite{Hees:2004} that
this also holds for heavy-light systems, and therefore could lead to
significantly faster thermalization for $c$ and $b$ quarks as compared
to perturbative QCD (pQCD) processes. Even at $T$=350\,MeV (which
roughly corresponds to initial temperatures at RHIC) and for
$\alpha_s$$=$0.4, elastic pQCD scattering, which is dominated by
$t$-channel gluon exchange, results in a thermal relaxation time for $c$
quarks above 10\,fm/$c$, well above a the typical QGP lifetime of
$\tau_{\mathrm{QGP}}$$\sim$5\,fm/$c$. Recent data on single-electron
spectra associated with semileptonic heavy-meson decays have
corroborated the need for nonperturbative HQ interactions in the QGP.
As shown in Fig.~\ref{fig_pqcd}, perturbative energy-loss calculations
appear insufficient to describe the strong suppression in the
$p_T$-spectra (left panel, using both elastic and radiative
interactions) and the large $v_2(p_T)$ (right panel, using radiative
energy loss with a substantially upscaled transport coefficient).
While the plots indicate a (lower) applicability limit of the pQCD
energy-loss approach of $p_T^e$$\simeq$2\,GeV, we will argue below that
nonperturbative effects may be relevant (or even dominant) up to at least
$p_T^e$$\simeq$5\,GeV. In particular, radiative energy-loss calculations
typically do not include the backward reactions (detailed balance) which
are essential for building up collective behavior of the heavy quarks in
the expanding thermal medium. As mentioned above, in this context a
Brownian-motion approach is well suited to address the thermalization of
the heavy quarks\cite{Hees:2004,Moore:2004,Svet88,Mustafa:200x}. Upon
expanding the Boltzmann equation in small momentum transfers, one
arrives at a Fokker-Planck equation for the HQ distribution function,
$f$,
\begin{equation}
  \frac{\partial f}{\partial t} = \gamma \frac{\partial (p f)}{\partial p}
  + D_p \frac{\partial^2 f}{\partial p^2} \ ,
 \end{equation}
 with (momentum) drag ($\gamma$) and diffusion ($D_p$) constants
 (related via $T=D_p/\gamma M_Q$). The latter are calculated from
 corresponding matrix elements for HQ scattering off light partons.
\begin{figure}[!t]
\begin{minipage}{5cm}
\vspace{-0.5cm}
\centering{\psfig{file=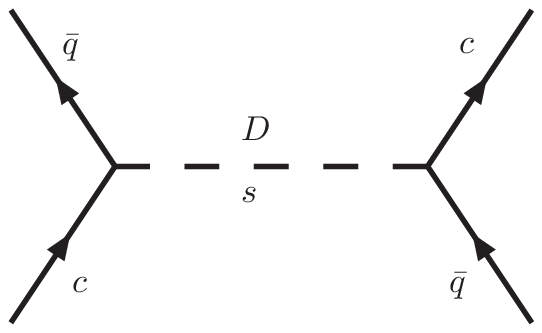,width=4.1cm}}\\
\centering{\psfig{file=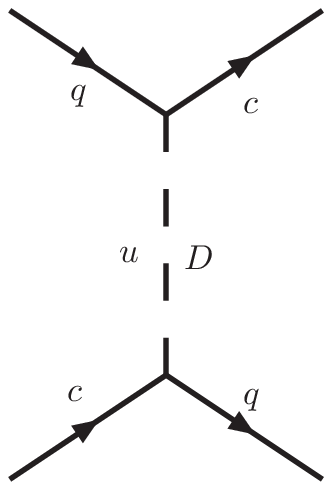,width=2.6cm}}
\end{minipage}
\begin{minipage}{6cm}
\vspace{0.5cm}
\psfig{file=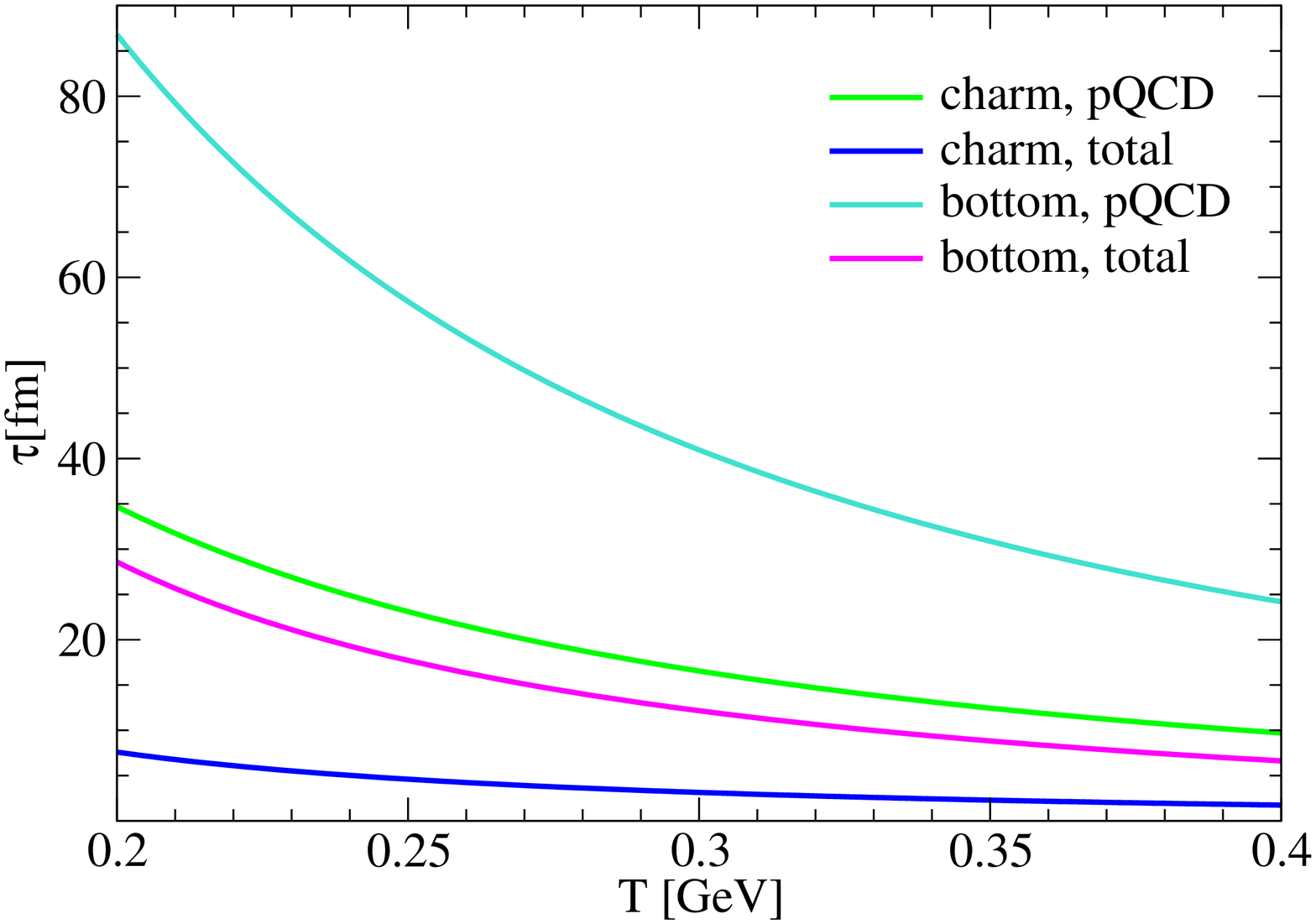,width=1.05\linewidth}
\end{minipage}
\caption{Elastic HQ-scattering processes via $D$-($B$-)meson like
  resonances (left panel) and resulting thermalization times compared to
  pQCD results (right panel)\protect\cite{Hees:2004}.}
\label{fig_tau}
\end{figure}
As mentioned above, our main ingredient here are resonance-mediated
elastic interactions in $s$- and $u$-channel (cf.~left panel of
Fig.~\ref{fig_tau}) which we model with an effective Lagrangian
according to\cite{Hees:2004}
\begin{equation}
\mathcal{L} = -G~\bar{Q}~\Gamma~\frac{1+\not{\!v}}{2}~\Phi~q + \mathrm{h.c} \ ;
\label{lag}
\end{equation}
$\Phi$ is the resonance field representing ``$D$'' and ``$B$'' mesons,
which we dress by evaluating the corresponding one-loop ($Q$-$\bar q$)
selfenergy. Parameters of the model are the resonance mass (which we fix
at 0.5\,GeV above the $Q$-$\bar{q}$ threshold) and coupling constant, $G$,
which determines the resonance width. HQ and chiral symmetry (above
$T_c$) imply degeneracy of the pseudo-/scalar and axial-/vector channels
represented by the Dirac matrices, $\Gamma$. The temperature dependence
of the resulting thermal relaxation times, $\tau_Q=\gamma^{-1}$, for
$c$- and $b$-quarks shows a factor of $\sim$3 reduction compared to
calculations where only pQCD elastic scattering is accounted for (right
panel of Fig.~\ref{fig_tau}). This is very significant for $c$-quarks as
$\tau_c$ is now comparable to (or even below) the duration of the QGP
phase at RHIC ($\tau_{\mathrm{QGP}}$$\sim$5\,fm/$c$), whereas $\tau_b$$>$$\tau_{\mathrm{QGP}}$
still, due to the large $b$-quark mass.

\begin{figure}[!t]
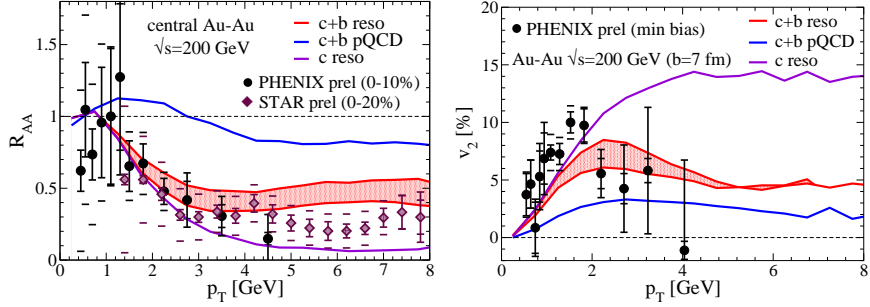

\vspace{-0.3cm}
\begin{minipage}{5.5cm}
\psfig{file=raa_e_cent-therm.eps,width=1.03\linewidth}
\end{minipage}
\hspace{0.2cm}
\begin{minipage}{5.5cm}
\psfig{file=v2_e_MB_therm-av-qm05data.eps,width=1.0\linewidth}
\end{minipage}
\caption{Calculations for single-electron spectra arising from
  semileptonic heavy-meson decays.  Left panel: Nuclear suppression
  factor right panel: elliptic flow.}
\label{fig_elec}
\end{figure}

The $T$- and $p$-dependent drag and diffusion coefficients have been
implemented into a relativistic Langevin simulation\cite{Hees:2004}
using an (elliptic) thermal fireball expansion for (semi-) central Au-Au
collisions at RHIC as a background medium, where the bulk flow is
adjusted to hydrodynamic simulations in accordance with experiment. At
the end of the QGP phase (which terminates in a mixed phase at
$T_c$$=$180\,MeV), the $c$- and $b$-quark output distributions from the
Langevin simulation are subjected to hadronization into $D$- and
$B$-mesons using the quark-coalescence model of Ref.~\cite{Greco:2003}.
Since the probability for coalescence is proportional to the light-quark
distribution functions, it preferentially occurs at lower $p_T$;
``left-over'' heavy quarks are hadronized via $\delta$-function
fragmentation. The extra contribution to momentum and $v_2$ from the
light quarks \emph{increases both} the $R_{AA}(p_T)$ and $v_2(p_T)$ of
the heavy-meson spectra, relative to a scheme with fragmentation only.
Rescattering in the hadronic phase has been neglected. The resulting
$D$- and $B$-meson spectra are decayed semileptonically resulting in a
nuclear suppression factor and elliptic flow which compare reasonably
well with recent RHIC data\cite{phenix-raa,star-raa,phenix-v2} up to
$p_T$$\simeq$5\,GeV (Fig.~\ref{fig_elec}).
At higher momenta, radiative energy loss (not included here) is expected
to contribute significantly\footnote{Note that in a chemically
  equilibrated QGP, the effects of induced gluon radiation are
  presumably smaller than in a pure gluon plasma as assumed in
  Refs.~\cite{Wicks:2005gt,Armesto:2005mz}; an estimate of (some) 3-body
  scattering diagrams has been performed in Ref.~\cite{Liu:2006xx}.}.

Another issue in interpreting the semileptonic $e^\pm$ spectra
concerns the relative contributions of charm and bottom.  Since bottom
quarks are much less affected by rescattering effects, their
contribution reduces the effects in both $R_{AA}$ and $v_2$ appreciably.
Following Refs.~\cite{Hees:2005,Rapp:2005at}, we have determined the
initial charm spectra by reproducing $D$- and $D^*$-spectra in d-Au
collisions\cite{star-D}, calculated the pertinent $e^\pm$ spectra and
assigned the missing yield relative to the data\cite{star-pp} to bottom
contributions. In this way, the crossing between charm and bottom
contribution in $p$-$p$ collisions occurs at $p_T$$\simeq$5\,GeV. Due to
the strong quenching of the $c$-quark spectra in the QGP this crossing
is shifted down to $p_T$$\simeq$2.5-3\,GeV in Au-Au collisions. For charm
contributions only, our nonperturbative rescattering mechanism results
in a pertinent electron $R_{AA}$ as low as 0.1 and an elliptic flow as
large as 14\%. Thus, the severeness of the ``$e^\pm$ puzzle'' partially
hinges on the baseline spectra in more elementary systems ($p$-$p$ and
d-Au). Obviously, a direct measurement of $D$-mesons in Au-Au would be
very valuable.

\section{Heavy Quarkonia in the QGP}
\label{sec_hidden}
\subsection{Potential Models, Spectral Functions and Lattice QCD}
\label{ssec_spec}
Heavy quarkonia ($c$-$\bar c$ and $b$-$\bar b$ bound states) have long
been identified as suitable objects to quantitatively investigate the
properties of QCD, including color confinement\cite{Novikov:1977dq}. A
particularly appealing feature is the applicability of potential models
as a nonrelativistic effective-theory approximation to QCD
(cf.~Ref.~\cite{Brambilla:2004wf} for a recent review). In the vacuum,
the heavy-quarkonium spectrum is indeed reasonably well described by the
``Cornell potential'', consisting of a Coulomb plus linear confining
term. Subsequently it was realized that, when immersing quarkonia into a
medium of deconfined color charges, Debye screening will reduce the
binding and eventually lead to the dissolution of the bound
state\cite{Matsui:1986dk,Karsch:1987pv}; already slightly above $T_c$, a
substantial (factor $\sim$2-3) reduction in charmonium and bottomonium
binding energies, $\varepsilon_B$, has been found, together with the
possibility that the ($S$-wave) ground states ($\eta_{c,b}$, $J/\psi$,
$\Upsilon(1S)$) survive (well) above $T_c$. Modern lQCD (-based)
calculations qualitatively support this picture: On the one hand,
directly extracted charmonium spectral functions (in quenched
approximation) indicate resonance peaks up to
$\sim$2$T_c$\cite{Asakawa:2003re,Datta:2003ww}. On the other hand, (both
quenched and unquenched) lQCD results for the (color-singlet) free
energy,
\begin{equation}
F_{Q\bar Q}(r;T)=U_{Q\bar Q}(r;T) -T S_{Q\bar Q}(r;T) \ , 
\end{equation}
have been implemented into potential models within a
Schr\"odinger\cite{Shuryak:2003ty,Wong:2004zr,Alberico:2005xw,Mocsy:2005qw}
or Lippmann-Schwinger equation\cite{Mannarelli:2005pz}. There is an
ongoing debate as to which quantity ($F_{Q\bar Q}$, the internal energy
$U_{Q\bar Q}$, or a linear combination thereof\cite{Wong:2006dz}) is the
most appropriate one to use as a potential. Maximal binding is obtained
with $U_{Q\bar Q}$, and even in this case, the ground-state charmonium
(bottomonium) binding energies are reduced to
$\varepsilon_B\simeq$0.1-0.2\,GeV (0.3-0.6\,GeV) at $T\simeq$1.5~$T_c$,
compared to $\sim$0.6\,GeV (1.1\,GeV) in vacuum. However, a quantitative
description of in-medium modifications of quarkonia requires at least
two additional components, i.e., the effects of parton-induced
dissociation reactions and of in-medium HQ masses.  The objective is
thus to construct an approach that (i) is consistent with finite-$T$
lQCD and (ii) allows for reliable applications to observables in
heavy-ion collisions. Only then the original idea\cite{Matsui:1986dk} of
using quarkonia as a tool to identify (and characterize) confinement (or
at least color screening) may be realized.

The interplay of medium effects becomes more transparent in terms of the
charmonium propagator, schematically written as
(nonrelativistic)\footnote{Writing Eq.~(\ref{Dpsi}) in this form implies
  that the $Q$-$\bar Q$ potential is strong enough to generate a pole in
  the scattering amplitude. A microscopic ($T$-matrix) approach is
  discussed below.}
\begin{equation}
D_\Psi(\omega;T)= \left[\omega-(2m_c^*-\varepsilon_B) 
+ {\rm i} \Gamma_\Psi \right]^{-1} \ .
\label{Dpsi}
\end{equation}
It shows that screening (affecting the binding energy and thus the real
part of the inverse propagator), and parton-induced dissociation
(governing the width, $\Gamma_\Psi$, of the spectral function,
Im~$D_\Psi$) are not mutually exclusive but have to be taken into
account simultaneously\cite{Ropke:1988zz}.

\begin{figure}[!t]
\vspace{-0.3cm}
\begin{minipage}{5.5cm}
\psfig{file=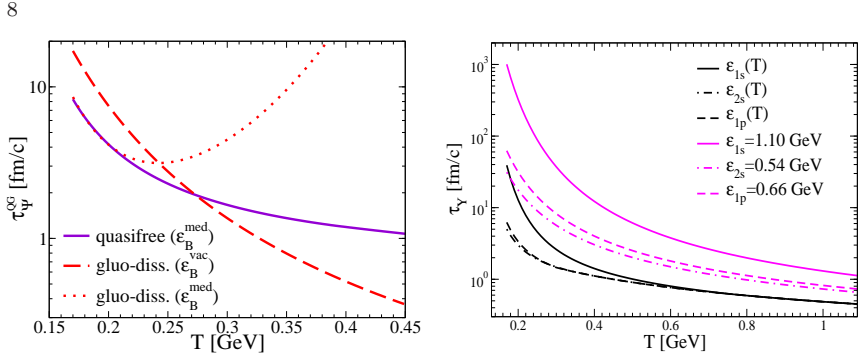,width=1.0\linewidth}
\end{minipage}
\hspace{0.1cm}
\begin{minipage}{5.5cm}
\psfig{file=tau_Y-qfree.eps,width=1.0\linewidth,height=0.75\linewidth}
\end{minipage}
\caption{Lifetimes of $J/\psi$ (left panel, 
    gluo-dissociation vs. quasifree dissociation
    mechanisms\protect\cite{Grandchamp:2001}) and $Y$ states (right panel,
    quasifree dissociation\protect\cite{Grandchamp:2005yw}) in the QGP.}
\label{fig_tau-onium}
\end{figure}

The inelastic width (or dissociation rate) can be expressed as
\begin{equation}
\Gamma_\Psi = (\tau_\Psi)^{-1} 
= \int \frac{d^3k}{(2\pi)^3} \ f^{p}(\omega_k,T)  \
v_{\mathrm{rel}} \ \sigma^{\mathrm{diss}}_\Psi(s) \ 
\label{gamma}
\end{equation}
in terms of (thermal) parton-distribution functions, $f^{p}$
($p$$=$$q,\bar{q},g$), and the parton-induced break-up cross section,
$\sigma^{\mathrm{diss}}_\Psi$. The latter has first been evaluated for
gluo-dissociation\cite{gluodiss}, $g+\Psi\to c+\bar c$, using Coulomb
wave-functions for the quarkonia, leading to a $T$-dependent $J/\psi$
lifetime as shown in the left panel of Fig.~\ref{fig_tau-onium}.
For the vacuum $J/\psi$ binding energy, $\varepsilon_B$$=$0.63\,GeV, the
lifetime is well behaved showing a rather steep decrease with $T$
(dashed line). However, if an in-medium binding energy is used as
following from screened (or lQCD) potentials, the lifetime is initially
reduced (dotted line), but if the binding energy drops below
$\sim$0.1\,GeV, the phase space in the gluo-dissociation reaction
strongly shrinks leading to an unphysical increase of the $J/\psi$
lifetime with $T$. This artifact is even more pronounced for the less
bound excited states ($\chi_c$, $\psi$'). This problem has been remedied
by introducing the ``quasifree'' dissociation
process\cite{Grandchamp:2001}, $p+\Psi\to p+c+\bar c$, where a parton
knocks out the $Q$ (or $\bar Q$) from the bound state. With reasonable
values for the strong coupling constant ($\alpha_s$$\simeq$0.25) the
resulting charmonium widths are in the $\sim$0.1\,GeV range for
$T$$\sim$1.5~$T_c$ (as relevant for RHIC) with a well-behaved
$T$-dependence. The order of magnitude of the inelastic width is easily
reproduced using a simplified estimate of Eq.~(\ref{gamma}) according to
$\Gamma_\Psi$$\approx$$n_p(T) \sigma_{\mathrm{diss}} v_{\mathrm{rel}}$;
with a parton density $n_p$$\simeq$10\,fm$^{-3}$ (at
$T$$\approx$0.25\,GeV), a cross section of 1\,mb and
$v_{\mathrm{rel}}$$=$1/2 one finds $\Gamma_\Psi$$\approx$0.1\,GeV.

When applied to bottomonia, the same arguments
apply\cite{Grandchamp:2005yw}; the right panel of
Fig.~\ref{fig_tau-onium} illustrates the impact of in-medium binding
energies on bottomonium dissociation: in the RHIC temperature regime,
screening may lead to up to a factor of $\sim$20 (!) reduction in the
ground-state $\Upsilon(1S)$ lifetime, bringing it down to $\sim$2-5fm/$c$ at
$T$$=$0.3-0.4\,GeV. This has important consequences for bottomonium
suppression at RHIC (as elaborated in Sec.~\ref{ssec_phen} below),
rendering it a sensitive probe of color
screening\cite{Grandchamp:2005yw}.
 
\begin{figure}[!t]
\vspace{-0.3cm}
\begin{minipage}{5.5cm}
\psfig{file=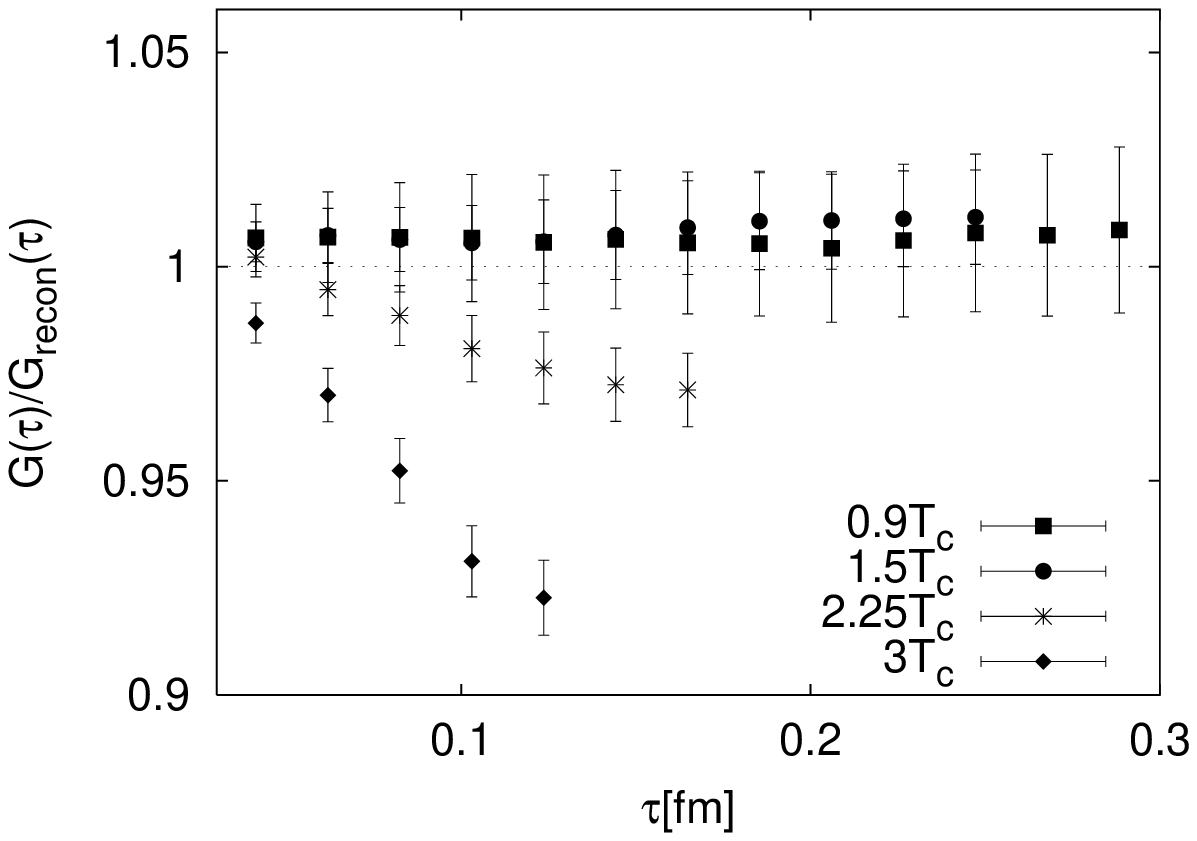,width=1.0\linewidth}
\end{minipage}
\hspace*{0.2cm}
\begin{minipage}{5.5cm}
\psfig{file=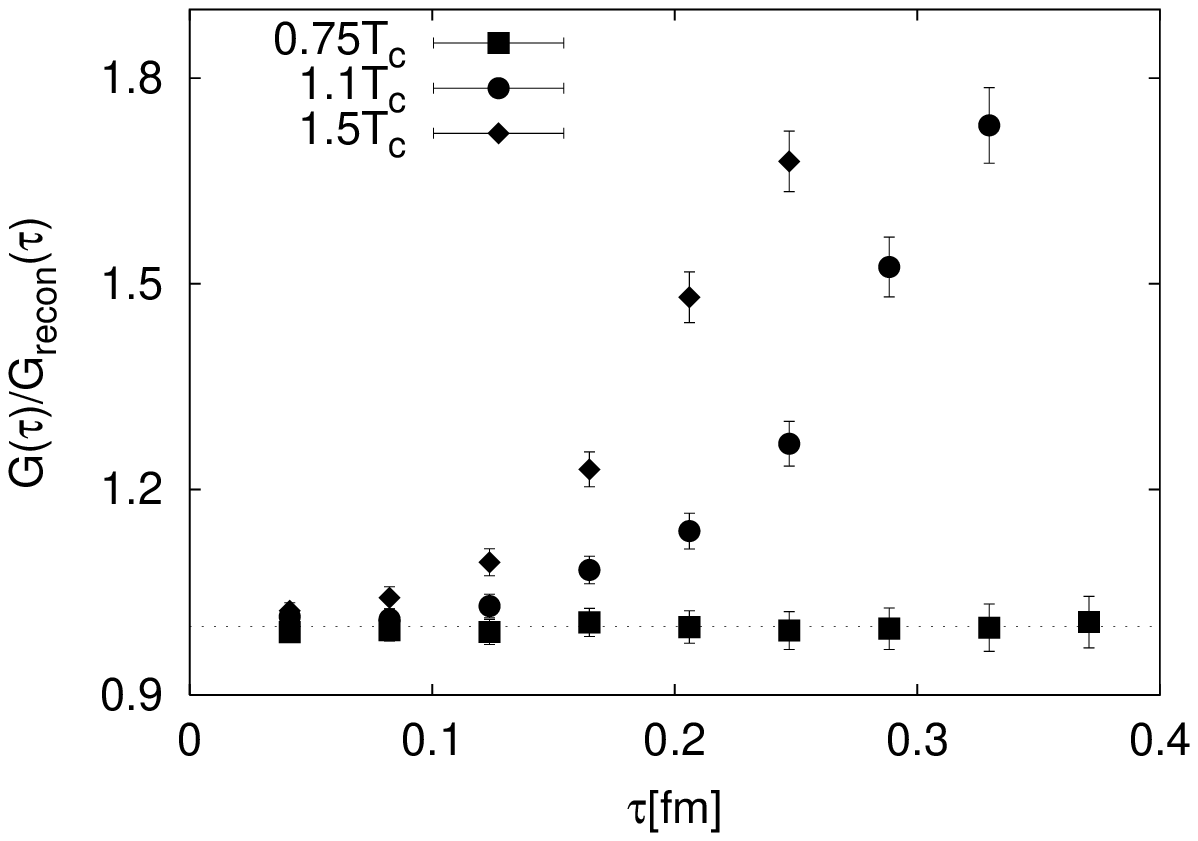,width=1.0\linewidth}
\end{minipage}
\caption{Finite-temperature euclidean correlation functions computed in
  lQCD in the pseudoscalar ($\eta_c$, left panel) and scalar ($\chi_c$,
  right panel) channels\protect\cite{Datta:2003ww}.  }
\label{fig_lat}
\end{figure}

A key issue for model descriptions of quarkonium spectral functions in
the QGP is the implementation of constraints from lQCD. The standard
object computed on the lattice is (the thermal expectation value of) a
hadronic current-current correlation function in euclidean time, $\tau$
(and at zero 3-momentum), in quantum-number channel,
$\alpha$\cite{Karsch:2003jg},
\begin{equation}
G_\alpha(\tau;T)=\langle j_\alpha(\tau) j^\dagger_\alpha(0)\rangle \ .  
\label{Ecor1}
\end{equation}
The connection to the spectral function, $\sigma_\alpha$, is given by a
convolution with a temperature kernel as
\begin{equation}
G_\alpha(\tau;T)= \int_0^\infty d\omega~\sigma_\alpha(\omega;T)
~\frac{\cosh[\omega(\tau-1/2T)]}{\sinh[\omega/2T]} \ .
\label{Ecor2}
\end{equation}
While euclidean correlators can be computed with good accuracy
(Fig.~\ref{fig_lat}),
their limited temporal extent at finite $T$, 0$\le$$\tau$$\le$1/$T$,
renders the extraction of (Minkowski) spectral functions more difficult.
On the contrary, model calculations of spectral functions are readily integrated
via Eq.~(\ref{Ecor2}) and compared to the rather precise euclidean
correlators from lQCD~\cite{Rapp:2002pn,Mocsy:2005qw}. To better exhibit
the $T$-dependence induced by the spectral function, both lattice and
model results for $G_\alpha(\tau;T)$ are commonly normalized to the
so-called ``reconstructed'' correlator, which follows from
Eq.~(\ref{Ecor2}) by inserting a free spectral function,
$\sigma_\alpha(\omega,T=0)$.

Eq.~(\ref{Ecor1}) requires the knowledge of the quarkonium spectral
function at all (positive) energies.  A schematic decomposition consists
of a bound-state part and a continuum,
\begin{equation}
\sigma_\alpha(\omega) = \sum_i 2 M_i F_i^2 \delta(\omega^2-M_i^2) +
\frac{3}{8\pi^2} \,\omega^2\,f_{\mathrm{cont}}(\omega,E_{\mathrm{thr}}) 
\Theta(\omega-E_{\mathrm{thr}}) \ , 
\end{equation}
where, for fixed $\alpha$, $i$ runs over all bound states with masses 
$M_i$ and couplings $F_i$, and the threshold energy ($E_{\mathrm{thr}}$) and
function $f_{\mathrm{cont}}$ characterize the onset and plateau of the continuum 
(usually taken from pQCD). 

In Ref.~\cite{Mocsy:2005qw}, $T$-dependent bound-state masses and
couplings have been evaluated by solving a Schr\"odinger equation with
in-medium potentials (screened Cornell or lQCD internal energy,
$U_{Q\bar Q}$); the threshold energy has been inferred from the
asymptotic ($r\to\infty$) value of the respective potential,
$E_{\mathrm{thr}}^{\mathrm{med}}(T)=2m_c+V(r\to\infty;T)$ (which
monotonously decreases for $T>T_c$). For the scalar ($\chi_c$) channel
it has been found that, despite a rather quickly dissolving bound-state
contribution, the correlator increases significantly (consistent with
the lattice result in Fig.~\ref{fig_lat}, right panel), due to the
decreasing continuum threshold. The latter also leads to an increasing
pseudoscalar correlator, which is not favored by lQCD
(Fig.~\ref{fig_lat}, left panel).

\begin{figure}[!t]
\begin{minipage}{5.5cm}
\psfig{file=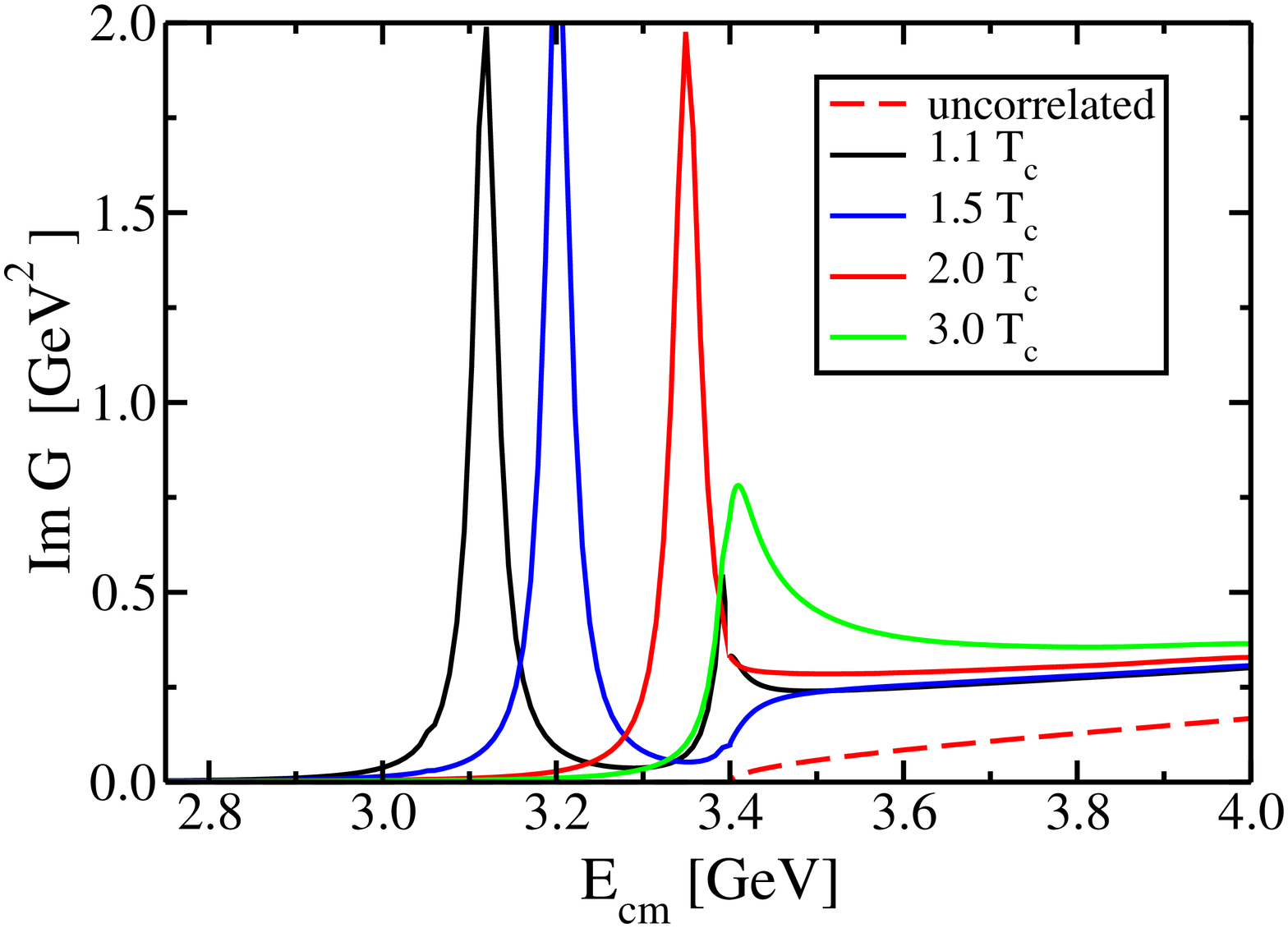,angle=-90,width=1.0\linewidth}
\end{minipage}
\hspace{0.2cm}
\begin{minipage}{5.5cm}
\psfig{file=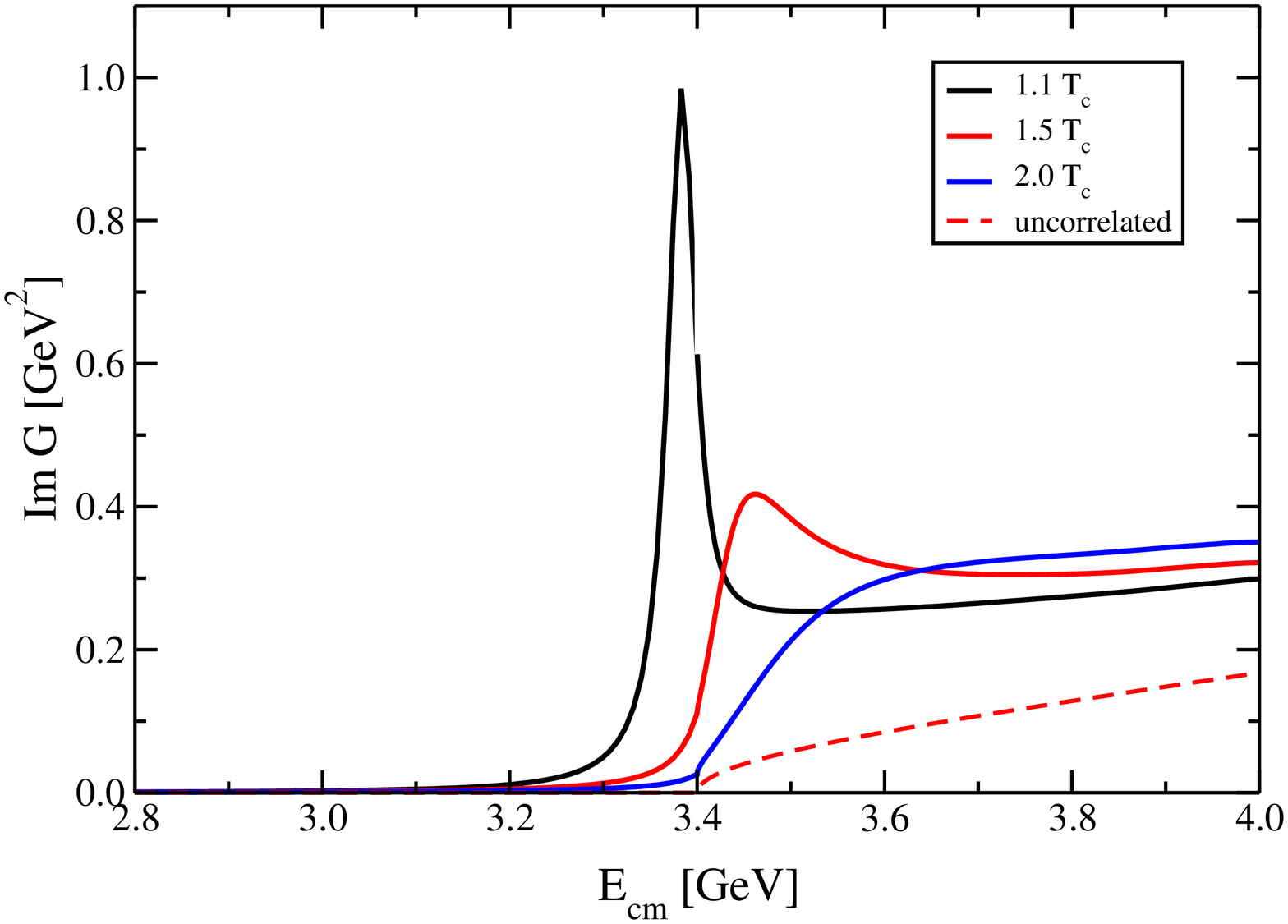,width=1.0\linewidth}
\end{minipage}
\caption{$S$- (left panel) and $P$-wave (right panel)
    charmonium-spectral functions from the $T$-matrix approach with an
    lQCD-internal energy as potential\protect\cite{Cabrera:2006xx}.}
\label{fig_spec}
\end{figure}

\begin{figure}[!t]
\vspace{-0.3cm}
\begin{minipage}{5.5cm}
\psfig{file=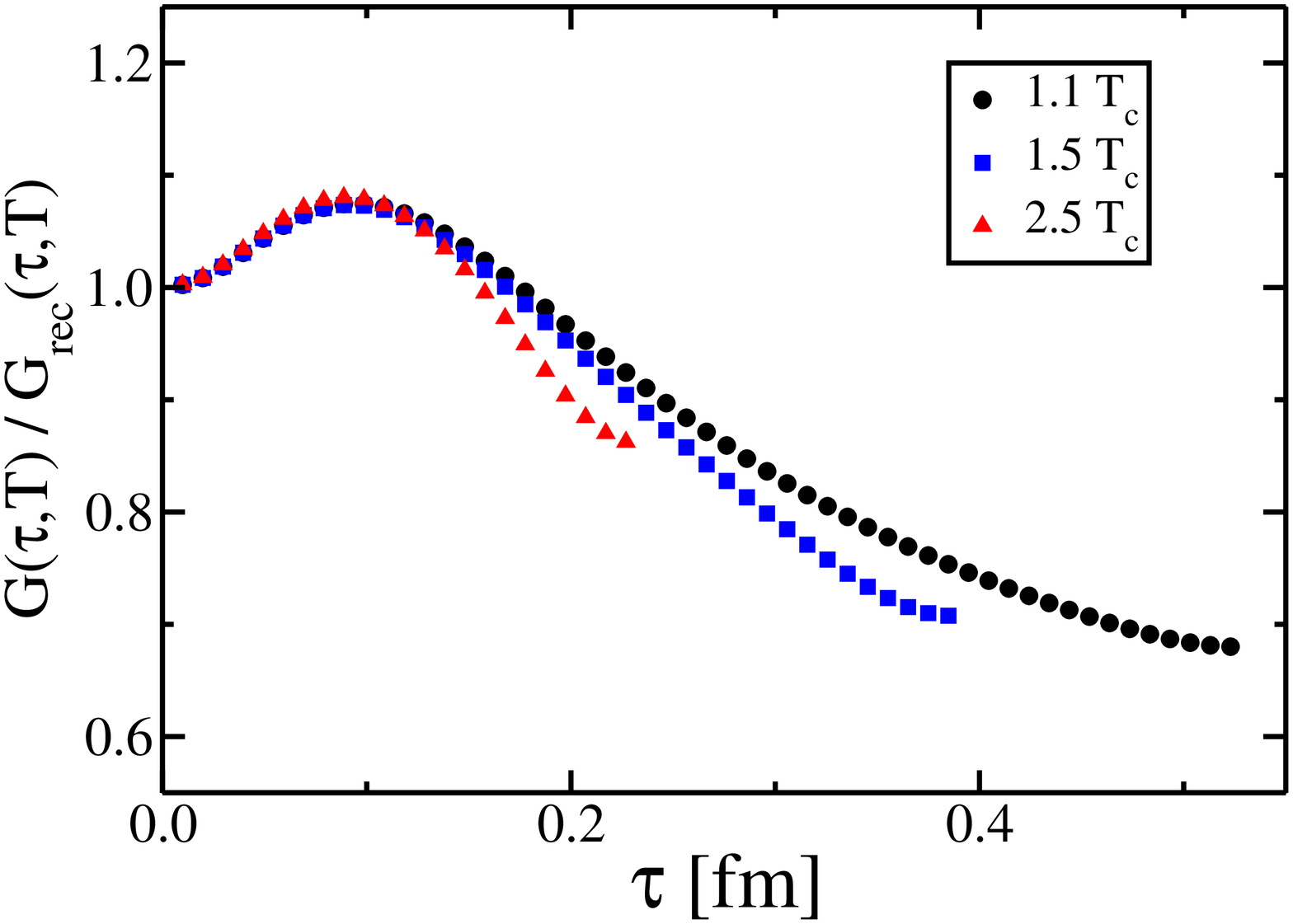,width=1.0\linewidth}
\end{minipage}
\hspace{0.2cm}
\begin{minipage}{5.5cm}
\psfig{file=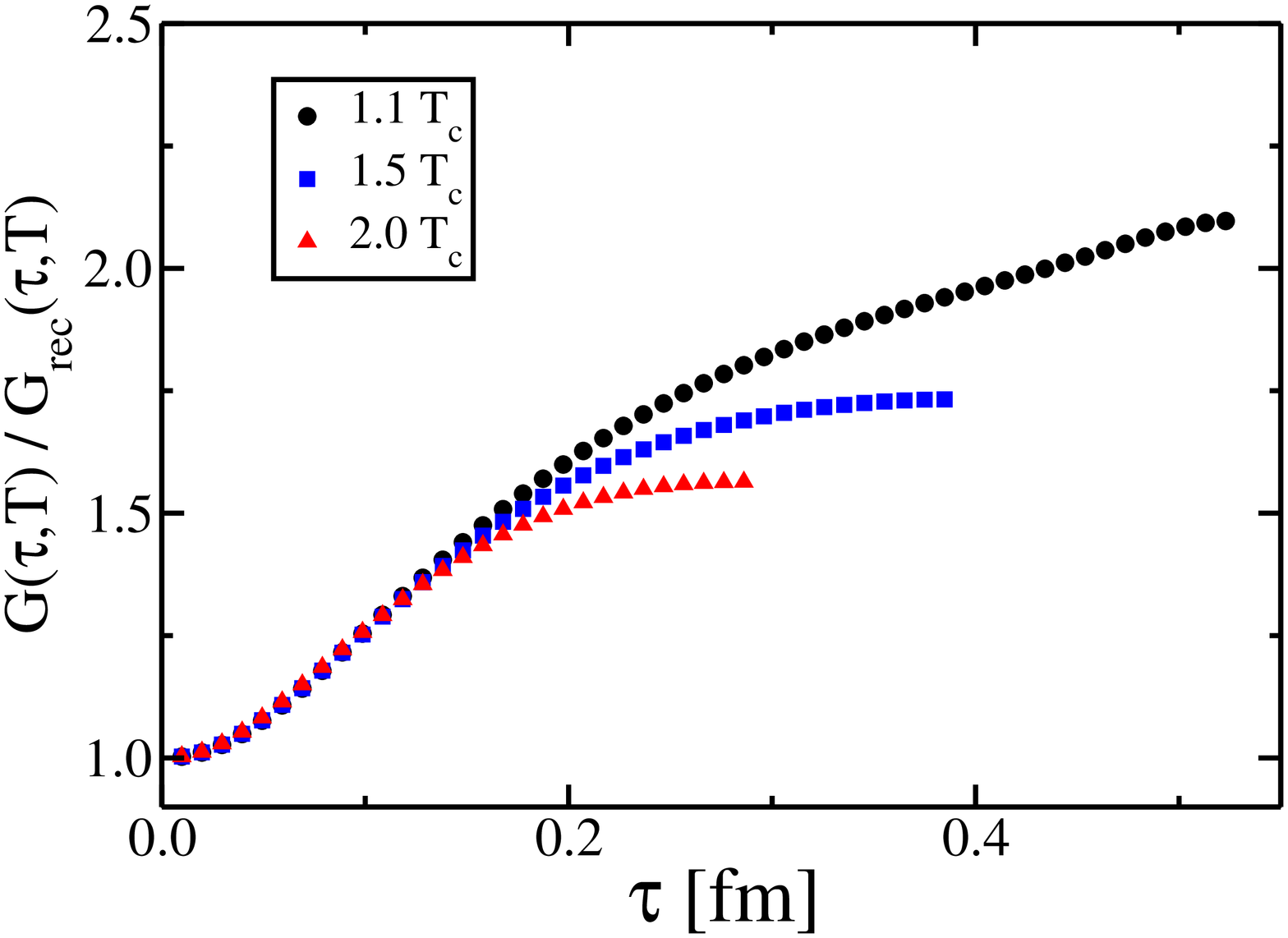,angle=-90,width=1.0\linewidth}
\end{minipage}
\caption{Normalized euclidean correlation function for $S$- (left panel)
    and $P$-wave (right panel) charmonium states within a $T$-matrix
    approach using lQCD-internal energies\protect\cite{Cabrera:2006xx}.}
\label{fig_eucl}
\end{figure}

In Ref.~\cite{Cabrera:2006xx} the separation of bound-state and
continuum parts is improved by employing a $T$-matrix
approach\cite{Celenza:1999tt,Mannarelli:2005pz} for the $Q$-$\bar Q$
interaction,
\begin{equation}
T_\alpha(E;q',q) = V_\alpha(q',q) - \frac{2}{\pi} \int_0^{\infty} 
dk\,k^2 \, V_\alpha(q',k)\, G_{\bar{Q}{Q}}(E;k) \, T_\alpha(E;k,q) 
 \ . 
\label{Tmat}
\end{equation}  
For $V_\alpha$ the Fourier transformed and partial-wave expanded lQCD
internal energy has been used. Spin-spin (hyperfine) interactions are
neglected implying degeneracy of states with fixed angular momentum
($S$-wave: $\eta_c$ and $J/\psi$, $P$-wave: $\chi_{c0,1,2}$).
$G_{\bar{Q}{Q}}(E;k)$ denotes the intermediate 2-particle propagator
including quark selfenergies. The correlation and spectral functions
follow from closing the external legs with 3-momentum integrations as
\begin{equation}
G_\alpha(E)=\int G_{\bar{Q}{Q}} + 
\int G_{\bar{Q}{Q}} \int T_\alpha \, G_{\bar{Q}{Q}} \ , \ 
\sigma_\alpha(E)= a \ {\rm Im}G_\alpha(E) \ , 
\end{equation}
where the coefficient $a$ depends on the channel $\alpha$.  For a fixed
$c$-quark mass of $m_c$=1.7\,GeV, the resulting charmonium spectral
functions confirm that bound states are supported in the $S$-wave up to
$\sim$3$T_c$ while dissolved in the $P$-wave below 1.5~$T_c$, see
Fig.~\ref{fig_spec}.
In both cases, however, one finds a large (nonperturbative) enhancement
of strength in the threshold region.  The corresponding euclidean
correlators are displayed in Fig.~\ref{fig_eucl}, normalized to a
reconstructed one using the vacuum bound-state spectrum and a
perturbative continuum with onset at the free $D\bar D$ threshold,
$E_{\mathrm{thr}}^{\mathrm{vac}}$=$2m_D$$\simeq$3.75\,GeV.  The $\chi_c$
correlator (right panel) is well above one, both due to a lower
in-medium threshold, $E_{\mathrm{thr}}^{\mathrm{med}}$$=$$2m_c$$=$3.4\,GeV,
and the nonperturbative rescattering enhancement. While this is
qualitatively consistent with lQCD, the temperature dependence is
opposite, indicative for a further threshold decrease with temperature.
In the $\eta_c$ channel (left panel), the significant reduction in
binding energy (due to color screening) reduces the correlator at large
$\tau$, qualitatively in line with lQCD. In fact, the latter shows less
reduction, leaving room for a $T$-dependent reduction of
$E_{\mathrm{thr}}^{\mathrm{med}}$ as well.

The scattering equation (\ref{Tmat}) is furthermore well suited to study
finite-width effects. E.g., when dressing the $c$-quarks with an
imaginary selfenergy corresponding to $\Gamma_c$$=$50~MeV, inducing a
charmonium width of $\Gamma_\Psi$$\simeq$100\,MeV (as in the left panel of
Fig.~\ref{fig_tau-onium}), the euclidean correlators vary by no more
than 2\%. This indicates that there is rather little sensitivity to
phenomenologically relevant values for the quarkonium widths.

\subsection{Quarkonium Phenomenology in Heavy-Ion Collisions}
\label{ssec_phen}
The formation of thermalized matter above $T_c$ in ultrarelativistic
heavy-ion collisions (recall the discussion in the introduction)
provides the basis for describing the production systematics of
quarkonia in terms of their in-medium properties as discussed in the
previous section. The suppression of the initially produced number of
quarkonia, $N_\Psi^i$, may be schematically written as proceeding
through 3 stages,
\begin{equation}
S_{\mathrm{tot}}= \exp(-\sigma_{\mathrm{nuc}}^{\mathrm{abs}}~\varrho_N~L) \  
   \exp(-\Gamma_{\mathrm{QGP}}~\tau_{\mathrm{QGP}}) \
   \exp(-\Gamma_{\mathrm{HG}}~\tau_{\mathrm{HG}}). 
\end{equation} 
With the suppression in the hadron gas (HG) believed to be small, and
the (``pre-equilibrium'') nuclear absorption (cross section) inferred
from $p$-$A$ experiments, the suppression factor directly probes the
dissociation rate of $\Psi$'s in the QGP. The observed
$J/\psi$-suppression pattern at SPS can indeed be well described using
the quasifree dissociation process with in-medium binding energies
consistent with lQCD-potential
models\cite{Grandchamp:2001,Grandchamp:2003uw}.

\begin{figure}[!t]
\begin{minipage}[t]{5.5cm}
\psfig{file=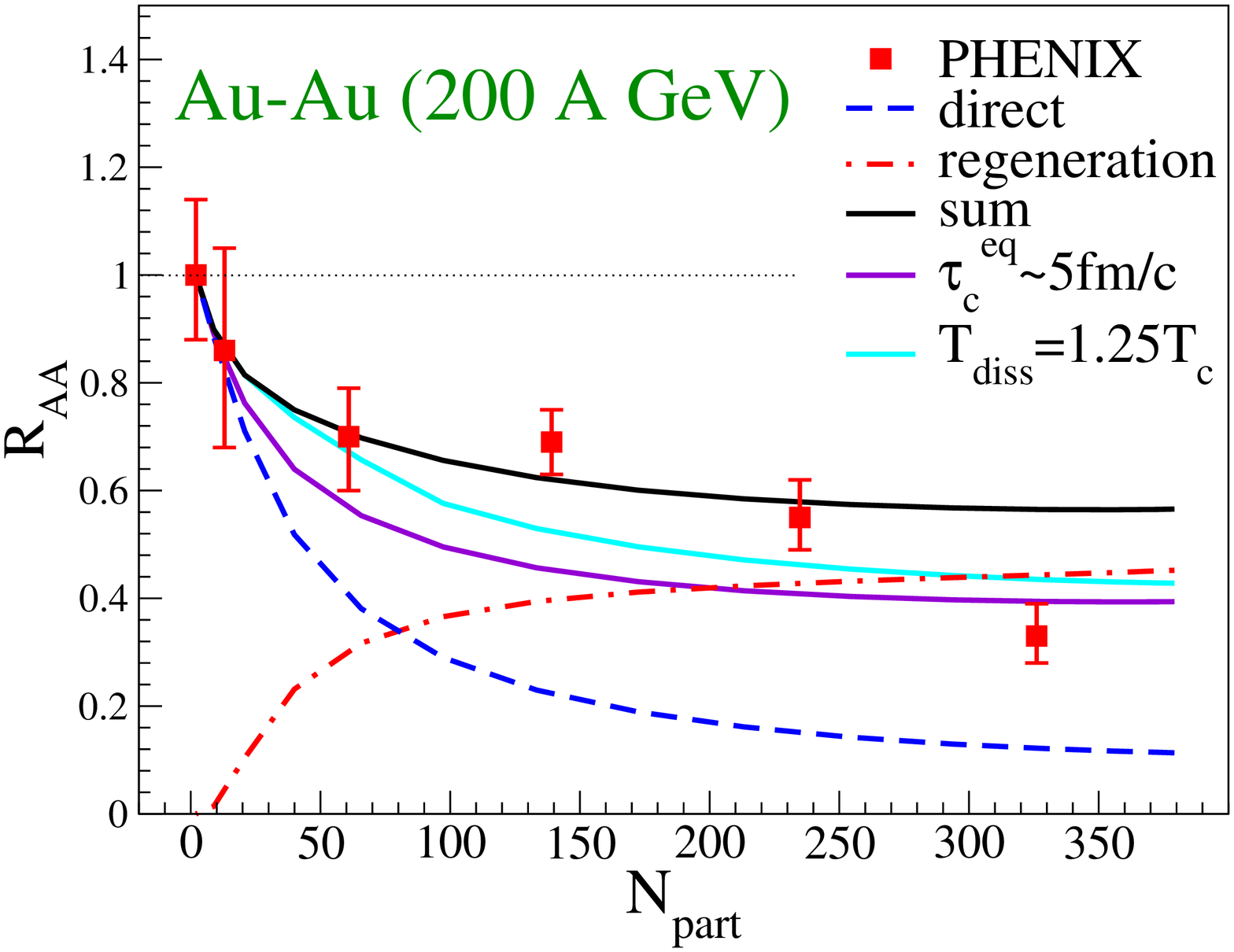,width=0.985\linewidth}
\end{minipage}
\hspace{0.2cm}
\begin{minipage}[t]{5.5cm}
\psfig{file=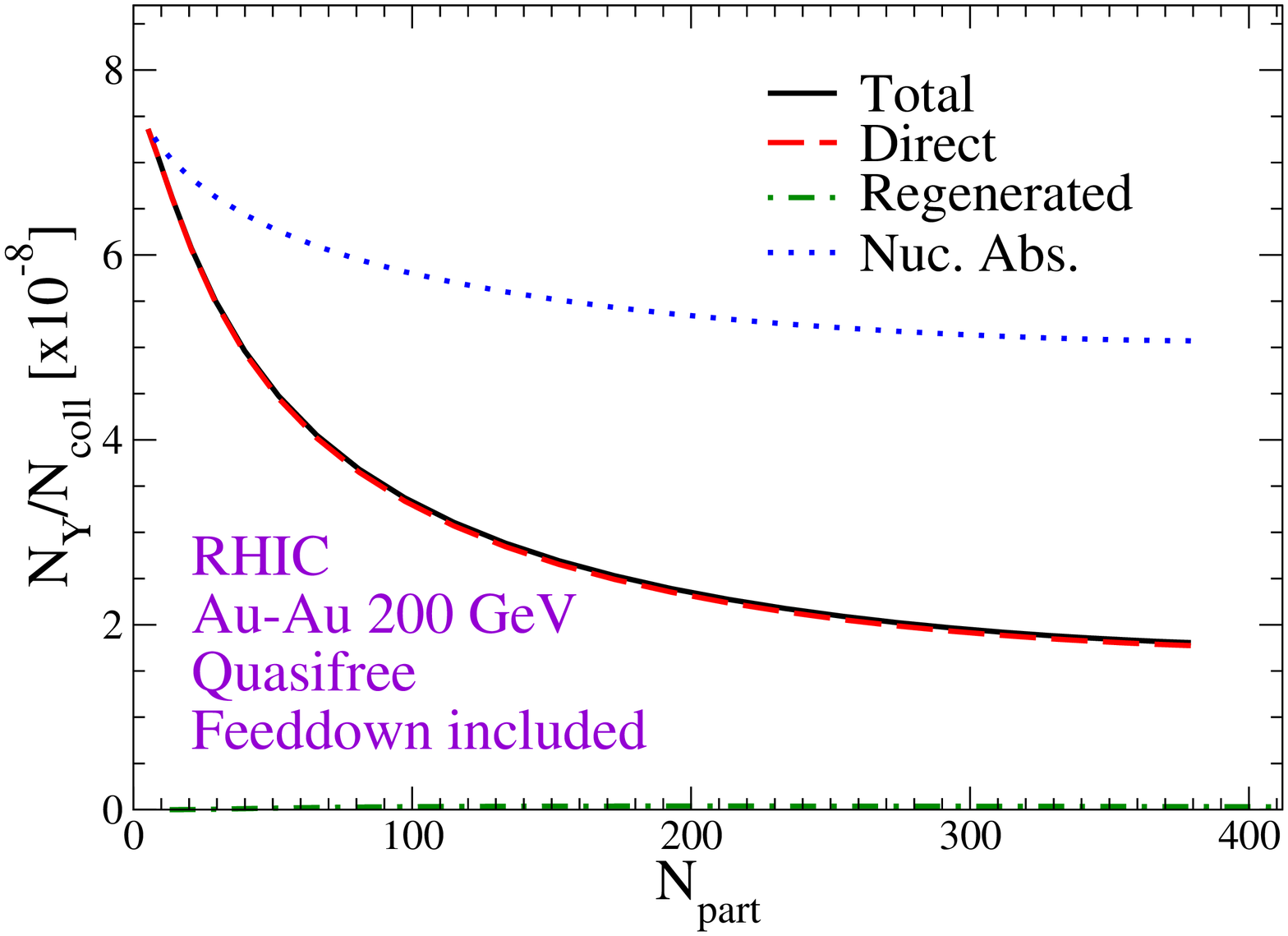,height=0.75\linewidth,width=0.9\linewidth}
\end{minipage}
\caption{Centrality dependence of the nuclear modification factor for
  $J/\psi$ (left panel)\protect\cite{Grandchamp:2003uw,Zhao:2006} and
  $\Upsilon$ (right panel)\protect\cite{Grandchamp:2005yw} at RHIC.}
\label{fig_rhic}
\end{figure}

At RHIC energy, copious production of $c\bar c$ pairs ($N_{c\bar
  c}$$\simeq$20 in central $\sqrt{s}$=200~AGeV Au-Au\cite{Adler:2004ta},
compared to 0.2 at SPS) opens the possibility of secondary charmonium
production\cite{pbm00,Grandchamp:2001}.  This is, after all, required by
detailed balance in the dissociation reaction,
$\Psi$+$g\rightleftharpoons c$+$\bar c$+$X$, provided the $\Psi$ state
still exists at the given
temperature\cite{Grandchamp:2003uw,Yan:2006ve}. If, in addition, the
$c$-quarks are close to thermal equilibrium, one can apply the rate
equation,
\begin{equation}
\frac{dN_\Psi}{dt} = - \Gamma_\Psi [N_\psi-N_\psi^{eq}(T)] \quad,
\end{equation}
for the time evolution of the $J/\psi$ number.  Pertinent predictions
(upper solid line in the left panel of Fig.~\ref{fig_rhic}) agree reasonably
well with current PHENIX data\cite{PereiraDaCosta:2005xz} suggesting
that the regeneration component becomes substantial in central Au-Au.
Corrections due to incomplete $c$-quark thermalization, as well as a
lower dissociation (and thus formation) temperature, reduce the
regeneration (lower solid curves in the left panel of
Fig.~\ref{fig_rhic}).
It has also been suggested\cite{Karsch:2005nk} that available $J/\psi$
data at SPS and RHIC are compatible with suppression of only $\chi_c$
and $\psi'$ (which make up $\sim$40\% of the inclusive $J/\psi$ yield in
$p$-$p$), with the direct $J/\psi$'s being unaffected.  However, if at
RHIC the $J/\psi$ width is of order 0.1\,GeV (cf.~previous section) and
the QGP lifetime at least 2fm/c, one finds a QGP suppression factor of
$S_{\mathrm{QGP}}$$\simeq$0.4.

Finally, turning to $\Upsilon$ production, the increase of the
$\Upsilon$ dissociation rate due to color screening (recall right panel
of Fig.~\ref{fig_tau-onium}) could lead to a $\Upsilon$ yield in central
Au-Au (right panel of Fig.~\ref{fig_rhic}) which is more suppressed for
the $J/\psi$ (left panel), and thus provide a rather striking QGP
signature\cite{Grandchamp:2005yw}.

\section{Conclusions}
\label{sec_concl}
Charm and bottom hadrons are valuable sensors of the sQGP. In the
open-flavor sector, the inadequacy of pQCD to account for the
suppression and collectivity of current $e^\pm$ spectra at RHIC may be a
rather direct indication of nonperturbative rescattering processes above
$T_c$. We have elaborated these in a concrete example using (elastic)
resonance interactions. The latter will have to be (i) evaluated more
microscopically, e.g., using input interactions from lQCD, and (ii)
combined with radiative energy loss.

In the quarkonium sector, steady progress is made in implementing
first-principle information from finite-$T$ lattice QCD into effective
models suitable for tests in heavy-ion reactions. $T$-matrix approaches
incorporate effects of color-screening, parton-induced dissociation, and
in-medium masses and widths of heavy quarks (which, in turn, connect to
the open-flavor sector). We have suggested an sQGP signature in terms of
stronger suppression for $\Upsilon$ relative to $J/\psi$, which would be
a direct proof of $J/\psi$ regeneration and carries large sensitivity to
color screening of bottomonia.

\section*{Acknowledgments}
This work was supported in part by a fellowship of the Spanish M.~E.~C.,
a Feodor-Lynen fellowship of the A.-v.-Humboldt foundation, and a U.S.
National Science Foundation CAREER Award under grant PHY-0449489.

    


\begin{thebibliography}{9}

\bibitem{phenix05}

K.~Adcox \textit{et al.} [PHENIX Collaboration], \textit{Nucl. Phys.}
{\bf A757}, 184 (2005). 

\bibitem{star05}
J.~Adams \textit{et al.} [STAR Collaboration], \textit{Nucl. Phys.} 
{\bf A757}, 102 (2005).

\bibitem{Gyulassy:2003mc}
  M.~Gyulassy, I.~Vitev, X.N.~Wang and B.W.~Zhang,
  arXiv:nucl-th/0302077.


\bibitem{phenix-raa}
  S.S.~Adler {\it et al.}  [PHENIX Collaboration],
\textit{Phys. Rev. Lett.} {\bf 96}, 032301 (2006).

\bibitem{star-raa}
  B.I.~Abelev {\it et al.} [STAR Collaboration],
  arXiv:nucl-ex/0607012.

\bibitem{Hees:2004}
H.~van Hees and R.~Rapp,  \textit{Phys. Rev. C} {\bf 71}, 034907 (2005).

\bibitem{Moore:2004}
G.D.~Moore and D.~Teaney,  \textit{Phys. Rev. C} {\bf 71}, 064904 (2005).

\bibitem{Mustafa:2005}
M.G.~Mustafa, \textit{Phys. Rev. D} {\bf 72}, 014905 (2005).

\bibitem{Hees:2005}
H.~van Hees, V.~Greco and R.~Rapp, \textit{Phys. Rev. C} {\bf 73}, 034913 (2006).

\bibitem{Wicks:2005gt}
  S.~Wicks {\it et al.}, 
  arXiv:nucl-th/0512076.

\bibitem{Svet88}
B.~Svetitsky \textit{Phys. Rev. D} {\bf 37}, 2484 (1988).

\bibitem{Armesto:2005mz}
N.~Armesto {\it et al.}, 
\textit{Phys. Lett. B} {\bf 637}, 362 (2006).


\bibitem{Mustafa:200x}
  D.~Pal and M.~G.~Mustafa,
  \textit{Phys. Rev. C} {\bf 60}, 034905 (1999).

\bibitem{Greco:2003}
V.~Greco, R.~Rapp and C.~M.~Ko \textit{Phys. Lett. B} {\bf 595}, 202 (2004).

\bibitem{phenix-v2}
 S.S.~Adler {\it et al.}  [PHENIX Collaboration],
  \textit{Phys. Rev. C} {\bf 72}, 024901 (2005);\\
  S.~Sakai,
  arXiv:nucl-ex/0510027.

\bibitem{Liu:2006xx}
  W.~Liu and C.~M.~Ko,
  arXiv:nucl-th/0603004.

\bibitem{Rapp:2005at}
  R.~Rapp, V.~Greco and H.~van Hees,
  arXiv:hep-ph/0510050.

\bibitem{star-D}
J.~Adams {\it et al.}  [STAR Collaboration], \textit{Phys. Rev.
Lett.} {\bf 94}, 062301 (2005).
                                                        
\bibitem{star-pp}
A.A.P.~Suaide {\it et al.} [STAR Collaboration] \textit{J. Phys. G} {\bf 30},
S1179 (2004).

\bibitem{Novikov:1977dq}
V.A.~Novikov {\it et al.}, 
\textit{Phys. Rept.} {\bf 41}, 1 (1978).

\bibitem{Brambilla:2004wf}

  N.~Brambilla {\it et al.},
  arXiv:hep-ph/0412158.

\bibitem{Matsui:1986dk}
  T.~Matsui and H.~Satz,
  \textit{Phys. Lett. B} {\bf 178}, 416 (1986).

\bibitem{Karsch:1987pv}
  F.~Karsch, M.T.~Mehr and H.~Satz,
  \textit{Z. Phys. C} {\bf 37}, 617 (1988).

\bibitem{Asakawa:2003re}
  M.~Asakawa and T.~Hatsuda,
  \textit{Phys. Rev. Lett.}  {\bf 92}, 012001 (2004).

\bibitem{Datta:2003ww}
S.~Datta {\it et al.}, 
\textit{Phys. Rev. D} {\bf 69}, 094507 (2004).

\bibitem{Shuryak:2003ty}
E.V.~Shuryak and I.~Zahed, \textit{Phys. Rev. C} {\bf 70}, 021901 (2004).

\bibitem{Wong:2004zr}
  C.Y.~Wong,
  \textit{Phys. Rev. C} {\bf 72}, 034906 (2005).

\bibitem{Alberico:2005xw}
  W.M.~Alberico {\it et al.}, 
  \textit{Phys. Rev. D} {\bf 72}, 114011 (2005).

\bibitem{Mocsy:2005qw}
  A.~Mocsy and P.~Petreczky,
  \textit{Phys. Rev. D} {\bf 73}, 074007 (2006).

\bibitem{Mannarelli:2005pz}
  M.~Mannarelli and R.~Rapp,
  \textit{Phys. Rev. C} {\bf 72}, 064905 (2005).

\bibitem{Wong:2006dz}
  C.Y.~Wong,
  arXiv:hep-ph/0606200.

\bibitem{Ropke:1988zz}
  G.~R{\"o}pke, D.~Blaschke and H.~Schulz,
  \textit{Phys. Rev. D} {\bf 38}, 3589 (1988).

\bibitem{gluodiss}
E.V.~Shuryak, \textit{Sov. J. Nucl. Phys.} \textbf{28}, 408 (1978); \\
G.~Bhanot and M.~Peskin, \textit{Nucl. Phys.} \textbf{B156}, 391 (1979).

\bibitem{Grandchamp:2001}
L.~Grandchamp and R.~Rapp, \textit{Phys. Lett. B} \textbf{523}, 60 (2001);
\textit{Nucl. Phys.} \textbf{A709}, 415 (2002).

\bibitem{Grandchamp:2005yw}
L.~Grandchamp {\it et al.}, 
\textit{Phys. Rev. C} {\bf 73}, 064906 (2006).

\bibitem{Karsch:2003jg}
F.~Karsch and E.~Laermann,
arXiv:hep-lat/0305025.


\bibitem{Rapp:2002pn}
R.~Rapp,
\textit{Eur. Phys. J. A} {\bf 18}, 459 (2003). 
                                      
\bibitem{Cabrera:2006xx}
D.~Cabrera and R.~Rapp, in preparation (2006).

\bibitem{Celenza:1999tt}
L.S.~Celenza, B.~Huang and C.M.~Shakin,
\textit{Phys. Rev. C} {\bf 59}, 1030 (1999).

\bibitem{Grandchamp:2003uw}
L.~Grandchamp {\it et al.}, 
\textit{Phys. Rev. Lett.}  {\bf 92}, 212301 (2004).

\bibitem{Adler:2004ta}
S.S.~Adler {\it et al.}  [PHENIX Collaboration],
\textit{Phys. Rev. Lett.}  {\bf 94}, 082301 (2005).

\bibitem{pbm00}
P.~Braun-Munzinger and J.~Stachel, \textit{Phys. Lett. B} {\bf 490}, 196
(2000).

\bibitem{Yan:2006ve}
  L.~Yan, P.~Zhuang and N.~Xu,
  arXiv:nucl-th/0608010.

\bibitem{PereiraDaCosta:2005xz}
  H.~Pereira Da Costa {\it et al.}~[PHENIX Collaboration],
  arXiv:nucl-ex/0510051.

\bibitem{Karsch:2005nk}
  F.~Karsch, D.~Kharzeev and H.~Satz,
  \textit{Phys. Lett. B} {\bf 637}, 75 (2006).

\bibitem{Zhao:2006}
X.~Zhao and R.~Rapp, work in progress (2006).

\end{thebibliography}
\end{document}